\documentclass{emulateapj}

\usepackage{graphicx}
\usepackage{ulem}
\usepackage[utf8]{inputenc}
\usepackage[T1]{fontenc}
\usepackage{url}
\usepackage{mdwlist}
\usepackage{amsmath} 
\usepackage{enumitem}

\makeatletter
\def\url@leostyle{%
 \@ifundefined{selectfont}{\def\UrlFont{\sf}}{\def\UrlFont{\small\ttfamily}}}
\makeatother
\begin{document}

\newcommand{\ls}{{_<\atop^{\sim}}}
\newcommand{\gs}{{_>\atop^{\sim}}}
\def \spose#1{\hbox  to 0pt{#1\hss}}  
\def \ls{\mathrel{\spose{\lower 3pt\hbox{$\sim$}}\raise  2.0pt\hbox{$<$}}}
\def \gs{\mathrel{\spose{\lower  3pt\hbox{$\sim$}}\raise 2.0pt\hbox{$>$}}}
\newcommand{\Ha}{\hbox{{\rm H}$\alpha$}}
\newcommand{\Hb}{\hbox{{\rm H}$\beta$}}
\newcommand{\Ovi}{\hbox{{\rm O}\kern 0.1em{\sc vi}}}
\newcommand{\OIII}{\hbox{[{\rm O}\kern 0.1em{\sc iii}]}}
\newcommand{\OII}{\hbox{[{\rm O}\kern 0.1em{\sc ii}]}}
\newcommand{\NII}{\hbox{[{\rm N}\kern 0.1em{\sc ii}]}}
\newcommand{\SII}{\hbox{[{\rm S}\kern 0.1em{\sc ii}]}}
\newcommand{\angstrom}{\textup{\AA}}
\newcommand\ionn[2]{#1$\;${\scshape{#2}}}

\font\btt=rm-lmtk10


\title{SDSS-IV M\lowercase{a}NGA: A serendipitous observation of a potential gas accretion event}

\shorttitle{A serendipitous observation of a potential gas accretion event}

\shortauthors{Cheung et al.}


\author{Edmond Cheung\altaffilmark{1}\dag, David V. Stark\altaffilmark{1}, Song Huang\altaffilmark{1}, Kate H. R. Rubin\altaffilmark{2}, Lihwai Lin\altaffilmark{3}, Christy Tremonti\altaffilmark{4}, Kai Zhang\altaffilmark{5}, Renbin Yan\altaffilmark{5}, Dmitry Bizyaev\altaffilmark{6,7}, M\'{e}d\'{e}ric Boquien\altaffilmark{8}, Joel R. Brownstein\altaffilmark{9}, Niv Drory\altaffilmark{10}, Joseph D. Gelfand\altaffilmark{11,12}, Johan H. Knapen\altaffilmark{13, 14}, Roberto Maiolino\altaffilmark{15,16}, Olena Malanushenko\altaffilmark{6}, Karen L. Masters\altaffilmark{17}, Michael R. Merrifield\altaffilmark{18}, Zach Pace\altaffilmark{4}, Kaike Pan\altaffilmark{6}, Rogemar A. Riffel\altaffilmark{19,20}, Alexandre Roman-Lopes\altaffilmark{21}, Wiphu Rujopakarn\altaffilmark{1,22}, Donald P. Schneider\altaffilmark{23,24}, John P. Stott\altaffilmark{25}, Daniel Thomas\altaffilmark{17}, and Anne-Marie Weijmans\altaffilmark{26}} 

\altaffiltext{1}{Kavli Institute for the Physics and Mathematics of the Universe (WPI), The University of Tokyo Institutes for Advanced Study, The University of Tokyo, Kashiwa, Chiba 277-8583, Japan}
\altaffiltext{2}{Harvard-Smithsonian Center for Astrophysics, 60 Garden Street, Cambridge, MA 02138, USA}
\altaffiltext{3}{Institute of Astronomy and Astrophysics, Academia Sinica, Taipei 106, Taiwan}
\altaffiltext{4}{Department of Astronomy, University of Wisconsin-Madison, 475 North Charter Street, Madison, WI 53706, USA}
\altaffiltext{5}{Department of Physics and Astronomy, University of Kentucky, 505 Rose Street, Lexington, KY 40506-0055, USA}
\altaffiltext{6}{Apache Point Observatory and New Mexico State University, P.O. Box 59, Sunspot, NM, 88349-0059, USA}
\altaffiltext{7}{Sternberg Astronomical Institute, Moscow State University, Moscow}
\altaffiltext{8}{Unidad de Astronom\'{i}a, Universidad de Antofagasta, Avenida Angamos 601, Antofagasta 1270300, Chile}
\altaffiltext{9}{Department of Physics and Astronomy, University of Utah, 115 S. 1400 E., Salt Lake City, UT 84112, USA}
\altaffiltext{10}{McDonald Observatory, Department of Astronomy, University of Texas at Austin, 1 University Station, Austin, TX 78712-0259, USA}
\altaffiltext{11}{NYU Abu Dhabi, P.O. Box 129188, Abu Dhabi, UAE}
\altaffiltext{12}{Center for Cosmology and Particle Physics, New York University, Meyer Hall of Physics, 4 Washington Place, New York, NY 10003, USA}
\altaffiltext{13}{Instituto de Astrof\'\i sica de Canarias, E-38205 La Laguna, Tenerife, Spain}
\altaffiltext{14}{Departamento de Astrof\'\i sica, Universidad de La Laguna, E-38205 La Laguna, Tenerife, Spain}
\altaffiltext{15}{Cavendish Laboratory, University of Cambridge 19 J. J. Thomson Avenue, Cambridge CB3 0HE, UK}
\altaffiltext{16}{Kavli Institute for Cosmology, University of Cambridge, Madingley Road, Cambridge CB3 0HA, UK}
\altaffiltext{17}{Institute for Cosmology and Gravitation, University of Portsmouth, Dennis Sciama Building, Burnaby Road, Portsmouth PO1 3FX}
\altaffiltext{18}{School of Physics and Astronomy, University of Nottingham, University Park, Nottingham, NG7 2RD, UK}
\altaffiltext{19}{Departamento de F\'{i}sica, Centro de Cie\^{n}cias Naturais e Exatas, Universidade Federal de Santa Maria, 97105-900 Santa Maria, RS, Brazil}
\altaffiltext{20}{Laborat\'orio Interinstitucional de e-Astronomia - LIneA, Rua Gal. Jos\'e Cristino 77, Rio de Janeiro, RJ - 20921-400, Brazil}
\altaffiltext{21}{Departamento de F\'{i}sica y Astronom\'{i}a, Facultad de Ciencias, Universidad de La Serena, Cisternas 1200, La Serena, Chile}
\altaffiltext{22}{Department of Physics, Faculty of Science, Chulalongkorn University, 254 Phayathai Road, Pathumwan, Bangkok 10330, Thailand}
\altaffiltext{23}{Department of Astronomy and Astrophysics, The Pennsylvania State University, University Park, PA 16802}
\altaffiltext{24}{Institute for Gravitation and the Cosmos, The Pennsylvania State University, University Park, PA 16802}
\altaffiltext{25}{Sub-department of Astrophysics, Department of Physics, University of Oxford, Denys Wilkinson Building, Keble Road, Oxford OX1 3RH, UK}
\altaffiltext{26}{School of Physics and Astronomy, University of St Andrews, North Haugh, St Andrews, Fife KY16 9SS, UK}

\altaffiltext{\dag}{ec2250@gmail.com}


\begin{abstract}

The nature of warm, ionized gas outside of galaxies may illuminate several key galaxy evolutionary processes. A serendipitous observation by the MaNGA survey has revealed a large, asymmetric $\Ha$ complex with no optical counterpart that extends $\approx8\arcsec$ ($\approx6.3$ kpc) beyond the effective radius of a dusty, starbursting galaxy. This $\Ha$ extension is approximately three times the effective radius of the host galaxy and displays a tail-like morphology. We analyze its gas-phase metallicities, gaseous kinematics, and emission-line ratios, and discuss whether this $\Ha$ extension could be diffuse ionized gas, a gas accretion event, or something else. We find that this warm, ionized gas structure is most consistent with gas accretion through recycled wind material, which could be an important process that regulates the low-mass end of the galaxy stellar mass function.

 \end{abstract}



\section{Introduction} \label{sec:introduction}

\setcounter{footnote}{26}

Understanding the warm, ionized gas outside of galaxies is a critical aspect of galaxy evolution. To study this gas, there have been two main probes: (1) observations of extraplanar ionized gas in edge-on galaxies \citep[e.g.,][]{rand90,  tullmann00, otte01, miller03, rossa03} and (2) low ionization metal-line absorption studies using quasar sightlines \citep[e.g.,][]{tumlinson11, werk13, werk14}. 

These studies have led to the discovery of diffuse ionized gas (DIG; \citealt{hoyle63, reynolds85}), which is a layer of warm, low-density ionized gas that extends out to several kpc into the haloes of galaxies, and the confirmation of the circumgalactic medium (CGM; \citealt{bergeron86, lanzetta95}), which is a gas reservoir containing warm, ionized gas that is of even lower density than the DIG and extends hundreds of kpc into the haloes of galaxies. But how this gas relates to the evolution of their host galaxies is an open question.

In this work, we further our understanding of warm, ionized gas in the haloes of galaxies by studying a rare and unusual gas complex in the SDSS-IV MaNGA survey \citep{bundy15}. Designed to observe galaxies out to a maximum radius of 2.5 effective radius ($R_{\rm e}$), the MaNGA survey has observed a dusty, starbursting\footnote{We define starbursts as galaxies with $\log~\Sigma_{\rm SFR} > -1 ~\rm M_{\odot}~yr^{-1}~kpc^{-2}$ \citep{kennicutt12}} galaxy that is on the upper end of the mass-metallicity relationship \citep{tremonti04} out to 6.3$R_{\rm e}$ through a fortuitous overestimation of $R_{\rm e}$\footnote{MaNGA uses $R_{\rm e}$ measurements from the NASA-Sloan Atlas, which estimated $R_{\rm e}=7.4\arcsec$ for this galaxy. After masking out the bright, nearby stars, we used GALFIT \citep{peng02} to fit a single S\'ersic model to this galaxy, yielding $R_{\rm e}=2.6\arcsec$; see \S2}. This galaxy shows no signs of interaction and displays a large $\Ha$ extension with no optical counterpart in the Sloan Digital Sky Survey Data Release 7 (SDSS DR7; \citealt{york00, abazajian09}). 

Throughout this work, we assume a flat cosmological model with $H_{0} = 70$ km s$^{-1}$ Mpc$^{-1}$,   $\Omega_{m} = 0.30$, and  $\Omega_{\Lambda} =0.70$.

\begin{deluxetable*}{cccccccccccc}
\tabletypesize{\scriptsize}
\tablewidth{0pt} 
\tablecaption{{\bf Galaxy properties}}
\tablehead{   
\colhead{MaNGA-ID} &
\colhead{Plate-IFU} &
\colhead{RA} &
\colhead{DEC} &
\colhead{$z$\tablenotemark{a}} &
\colhead{$\log~M_*$\tablenotemark{b}} &
\colhead{$u-r$\tablenotemark{c}} &
\colhead{$\log~SFR$\tablenotemark{d}} &
\colhead{$\log~\Sigma_{\rm SFR}$\tablenotemark{e}} &
\colhead{$R_{\rm e}$\tablenotemark{f}} &
\colhead{$R_{\rm e}$} &
\colhead{$n$\tablenotemark{g}}  \\
\colhead{} &
\colhead{} &
\colhead{(J2000.0 deg)} &
\colhead{(J2000.0 deg)} &
\colhead{} &
\colhead{(M$_{\odot}$)} &
\colhead{} &
\colhead{(M$_{\odot}~\rm yr^{-1}$)} &
\colhead{(M$_{\odot}~\rm yr^{-1} kpc^{-2}$) } &
\colhead{($\arcsec$)} &
\colhead{(kpc)} &
\colhead{} 
} 
\startdata 
1-113700  &  8618-12703 & 319.45182 & 11.66059 & 0.038 & 9.77 &  2.05  & 0.22 & $-0.37$ & 2.6 & 2.0 & 3.7 
\enddata
\tablenotetext{a}{Spectroscopic redshift from NSA catalog.}
\tablenotetext{b}{Galaxy stellar mass from MPA-JHU DR7 data release.}
\tablenotetext{c}{Rest-frame $u-r$ color from NSA catalog.}
\tablenotetext{d}{Fiber star-formation rate from MPA-JHU DR7 data release.}
\tablenotetext{e}{Fiber star-formation rate surface density using the MPA-JHU DR7 data release.}
\tablenotetext{f}{Effective radius from GALFIT \citep{peng02}.}
\tablenotetext{g}{Galaxy S\'ersic index from GALFIT \citep{peng02}.}
\label{tab:properties}
\end{deluxetable*}

\begin{figure*}[t!] 
\centering
\includegraphics[scale=.7]{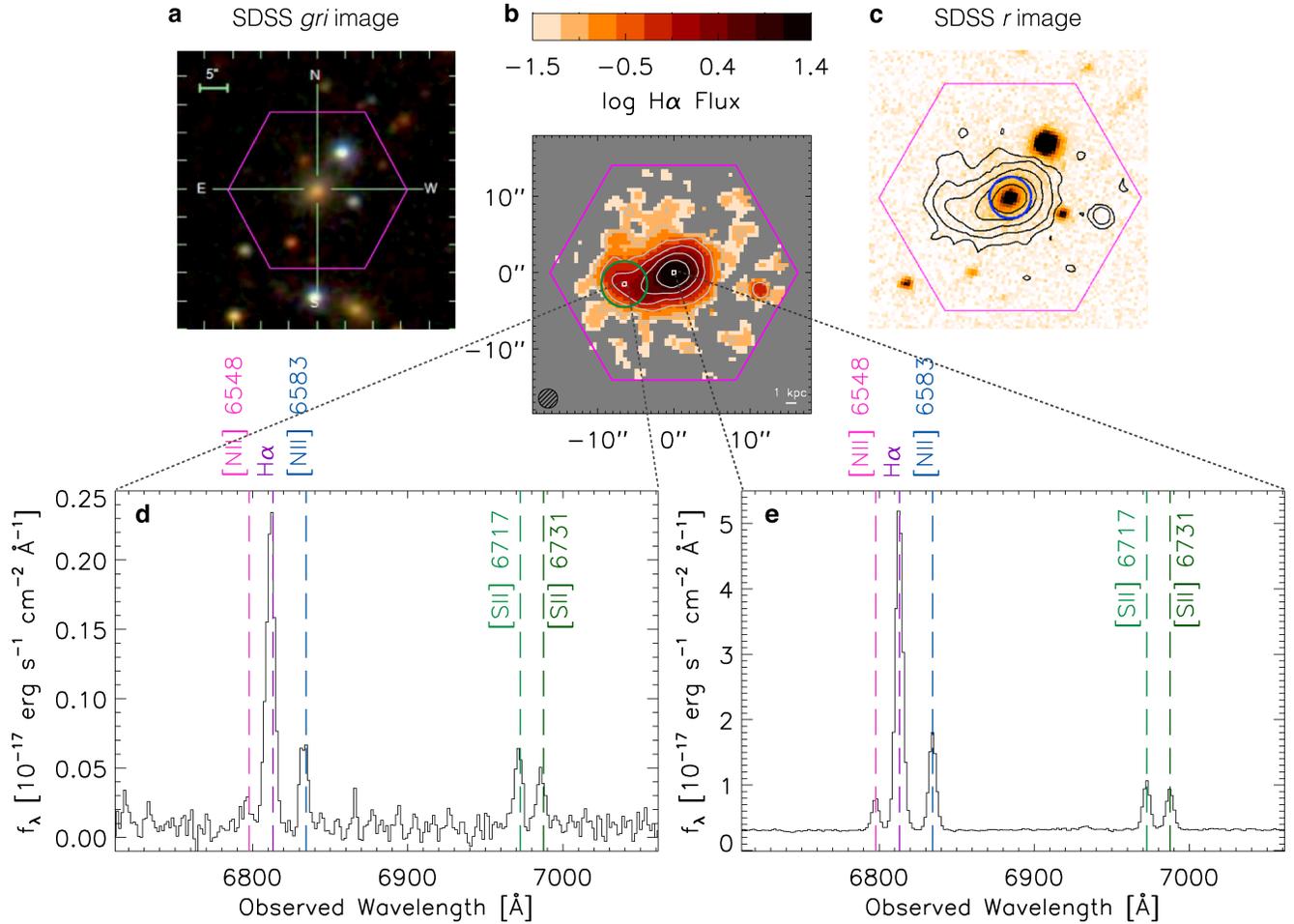}
\caption{ {\it a:} SDSS $gri$ color image of the object, with the MaNGA field of view in magenta. {\it b:} $\Ha$ flux map with contours of $\log~{\Ha}~\rm Flux =-1,~ -0.5, ~ 0, ~ 0.5, ~ \rm and ~ 1$; the flux units are $10^{-17} ~ \rm ergs ~ s^{-1} cm^{-2}$. There is an asymmetric extension in the $\Ha$ flux distribution to the left (East) of the host galaxy. The green circle, which has a radius of $3\arcsec$, is an approximation of this $\Ha$ extension---we refer to this as the ``$\Ha$ circle'' throughout the text. The lower left hatched circle represents the effective spatial resolution of MaNGA, FWHM $= 2.5\arcsec$. {\it c:} SDSS $r$ band image with the $\Ha$ flux contours superimposed; the blue circle marks the $R_{\rm e}$ of the host galaxy. There is no optical source in the region of the $\Ha$ extension. {\it d-e:} Spectra (before stellar continuum subtraction) from the indicated spaxels in the $\Ha$ extension {\it (d)} and in the center of the galaxy {\it (e)}. The vertical, dashed lines indicate the expected wavelengths of the $\NII$ doublet, $\Ha$, and $\SII$ doublet emission lines at the systemic velocity of the host galaxy.
\label{fig:spectra}}
\end{figure*} 

\section{Data} \label{sec:data}

\begin{figure*}[t!] 
\centering
\includegraphics[scale=1]{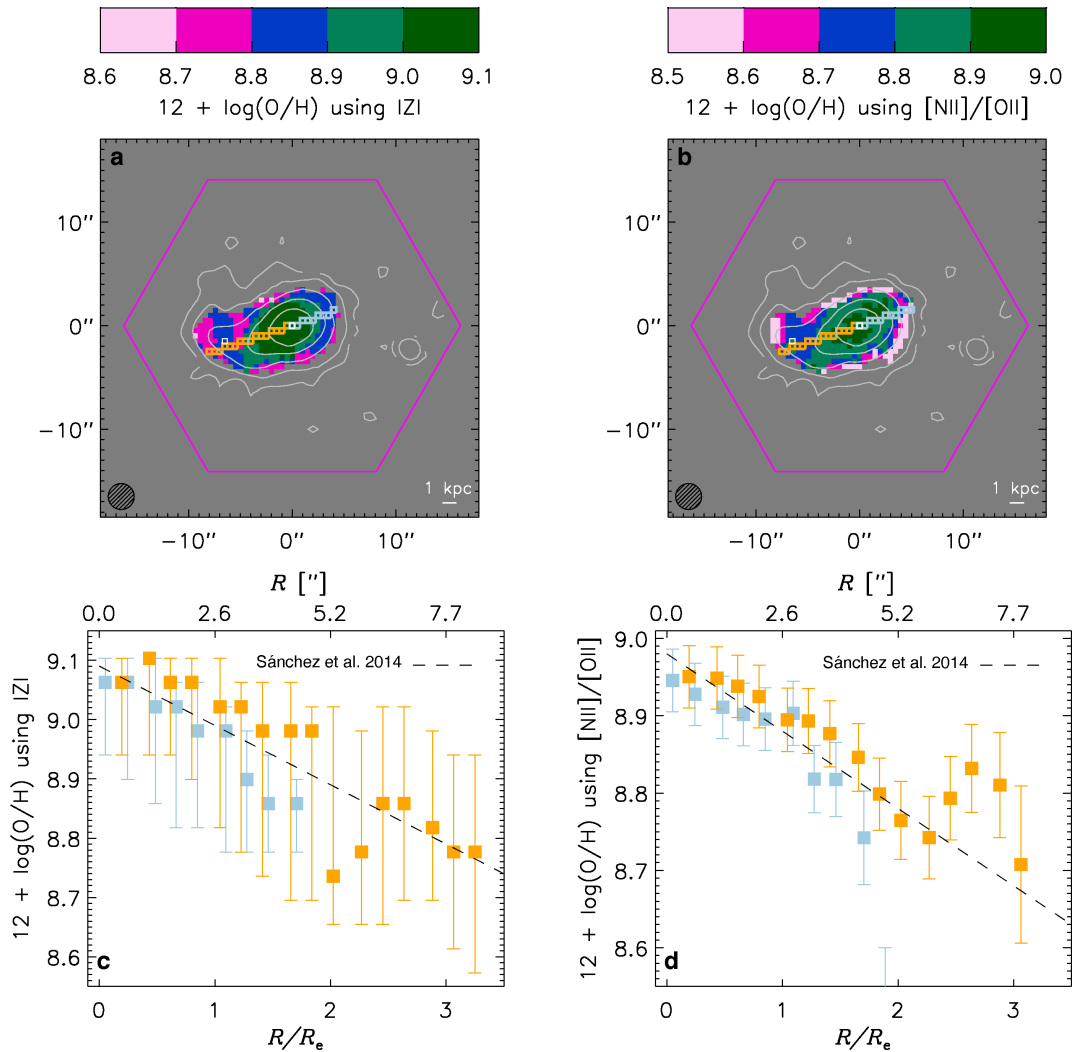}
\caption{The gaseous metallicity, 12+log(O/H), map of the system using {\tt IZI} ({\it a}) and the $\NII/\OII$ calibration ({\it b}). The $\Ha$ extension has lower gaseous metallicities, by $\approx0.25$ dex and $\approx0.15$ dex, respectively, than the center of the host galaxy at greater than $99.7\%$ confidence. The 12+log(O/H) profile of the highlighted spaxels in panels {\it a} and {\it b} using {\tt IZI} ({\it c}) and the $\NII/\OII$ calibration ({\it d}); the colors correspond to the colors of the highlighted spaxels. The characteristic metallicity gradient of non-interacting disks from \cite{sanchez14} is overlaid in the dashed line (with an arbitrary zero-point) and the error bars represent the 1$\sigma$ uncertainties. 
\label{fig:logoh}}
\end{figure*} 

The data used in this work are from the ongoing SDSS-IV MaNGA survey \citep{bundy15, drory15, law15, yan16, johnson16}, which is an integral field unit (IFU) survey that is taking resolved spectroscopy of 10,000 nearby galaxies with $\log~M_*/\rm M_{\odot}\gs9$. The survey uses the SDSS 2.5-meter telescope \citep{gunn06} and BOSS spectrographs \citep{smee13}; pilot studies using P-MaNGA data include \cite{li15, wilkinson15, belfiore15}. The $r$-band signal-to-noise $(S/N)$ in the galaxy outskirts is 4-8 \AA$^{-1}$~and the wavelength coverage is 3,600-10,000 \AA. The effective spatial resolution is $2.5''$ (full width at half maximum; FWHM), with an instrumental resolution of $\sigma\approx60$ km s$^{-1}$ and a spectral resolution of $R\sim2000$.

The MaNGA sample and data products were drawn from the internal MaNGA Product Launch-4 (MPL-4), which contains $1,368$ galaxies. Ancillary data are from the MPA-JHU DR7 value-added catalog\footnote{\href{http://www.mpa-garching.mpg.de/SDSS/DR7/}{http://www.mpa-garching.mpg.de/SDSS/DR7/}} and the NASA-Sloan Atlas\footnote{\href{http://www.nsatlas.org}{http://www.nsatlas.org}}. 

The stellar masses from the MPA-JHU DR7 value-added catalog are estimated by fitting a large grid of stellar population models from \cite{bruzual03} to the {\it ugriz} SDSS photometry, following the philosophy of \cite{kauffmann03} and \cite{salim07}. The fiber star formation rates from the MPA-JHU DR7 value-added catalog are estimated using the technique described in \cite{brinchmann04}, where galaxies with emission lines are fitted with the models of \cite{charlot01}. The fiber star formation rate surface density is calculated over the $3''$ diameter SDSS fiber.

The stellar continuum of each spaxel is fit by the MaNGA Data Analysis Pipeline (DAP; Westfall et al., in prep), which uses {\tt pPXF} \citep{cappellari04} and the MIUSCAT stellar population models \citep{vazdekis12}. Although the MIUSCAT templates are built to reflect simple stellar populations, the mix of templates used to construct the best-fit stellar continuum does not represent a physically-motivated stellar population.

Emission line fluxes are measured through simple flux-summing after subtraction of the stellar continuum (where there is stellar continuum). The wavelength passbands over which they are integrated are similar to that of \cite{yan06}, ranging from 10 \AA~to 20 \AA~around the central wavelength. For spectra that do not have measured stellar continua, we subtract a baseline continuum that is based on a linear fit to the red and blue sidebands; these sidebands extend approximately 100 \AA~beyond the central passbands, and have been chosen to avoid other strong emission lines.
 
We adopt the non-parametric quantities $v_{\rm peak}$ and $W_{80}$ to characterize the center (with respect to the systemic velocity of the host galaxy) and width (that contains $80\%$ of the flux) of the $\Ha$ emission line \citep[e.g.,][]{harrison14}. We only measure $v_{\rm peak}$ and $W_{80}$ in spaxels where the $\Ha$ flux has $S/N>5$. 

To estimate the gaseous metallicities, 12+log(O/H), we use the {\tt IZI} code \citep{blanc15}, which estimates ionization parameter and gaseous metallicity based on a bayesian analysis of a model grid with a set of input line fluxes; we used the \cite{dopita13} model grids, but obtain similar results with that of \cite{levesque10}. We provide \OII~3727,3729, \Hb, \OIII~5007, \Ha, \NII~6583, and \SII~6717,6731; all provided line fluxes have $S/N>3$. Throughout this work, we correct for reddening using the Balmer decrement and the \cite{fitzpatrick99} extinction law. The typical uncertainties of these {\tt IZI} metallicities are $0.1$-$0.3$ dex. 

We also estimate the gaseous metallicity using the \NII/\OII~calibration from \cite{kewley02}, which is relatively insensitive to variations in ionization parameter and is least sensitive to diffuse ionized gas contamination (Zhang et al. 2016, submitted). We estimate the uncertainties by adding in quadrature the intrinsic scatter of this calibration ($0.04$ dex; see \citealt{kewley02}) and the measurement error of $\log~\NII/\OII$ propagated through the calibration, resulting in typical uncertainties of $0.05$-$0.15$ dex. 

\section{Results} \label{sec:results}

Our main result is presented in Fig.~\ref{fig:spectra}. The SDSS $gri$ color image of the system is shown in Fig.~\ref{fig:spectra}a, with the MaNGA footprint outlined by the magenta hexagon. Fig.~\ref{fig:spectra}b displays the $\Ha$ flux map of this system, with the $\Ha$ flux contours in light grey. There is a large extension in the $\Ha$ flux distribution that extends $\approx8\arcsec$ ($\approx6.3$ kpc) to the left (East) beyond the effective radius of the host galaxy, and does not correspond to any optical source in SDSS. This striking feature is elucidated in Fig.~\ref{fig:spectra}c, where we present the $\Ha$ flux contours superimposed on the SDSS $r$ band image, with the blue circle marking the $R_{\rm e}$ of the host galaxy.

\begin{figure*}[t!] 
\centering
\includegraphics[scale=1]{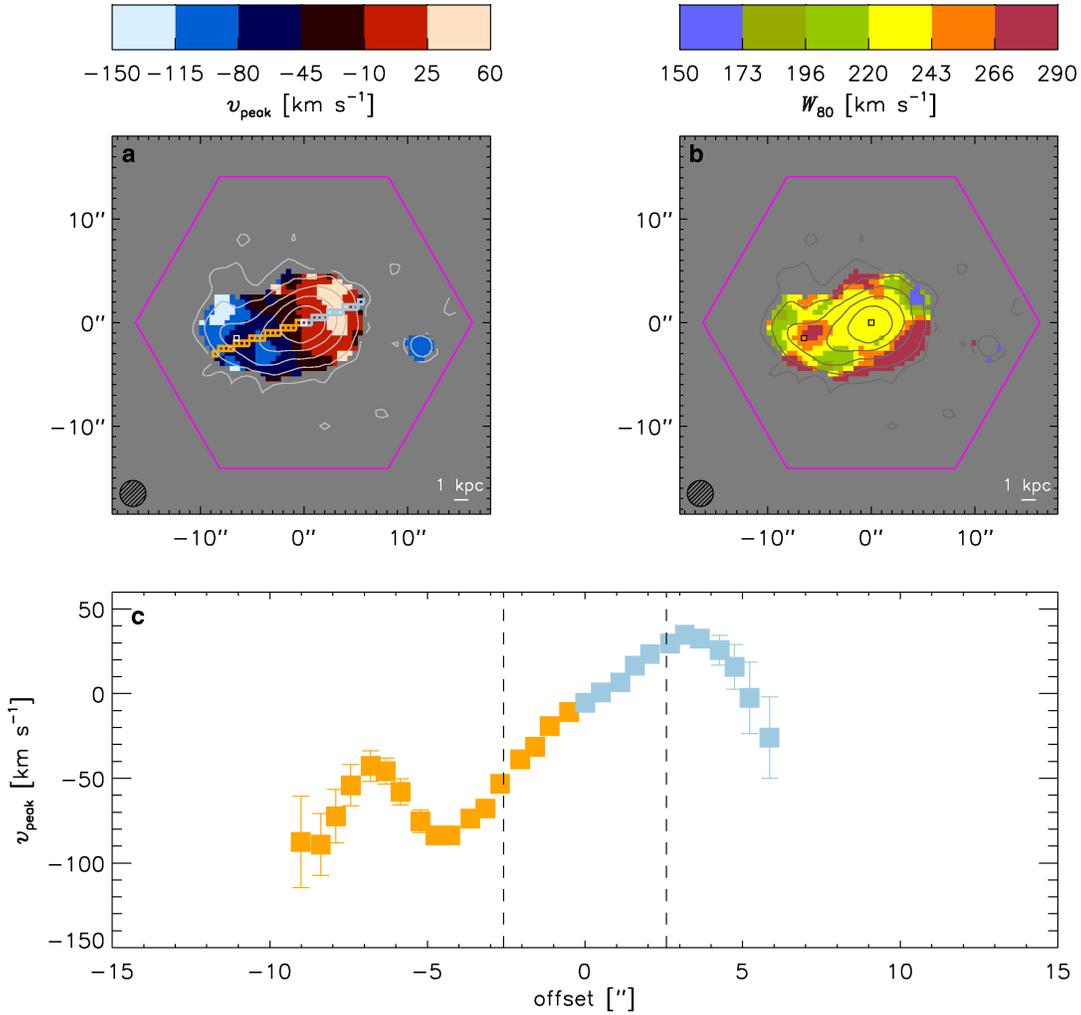}
\caption{ {\it a:} Peak velocity, $v_{\rm peak}$, of the system. {\it b:} Line width, $W_{\rm 80}$, of the system. {\it c:} Velocity profile of the highlighted spaxels in {\it a}, where the color corresponds to the color of the highlighted spaxels. The dashed, vertical lines mark $R_{\rm e}$ of the host galaxy and the error bars are the 1$\sigma$ uncertainties estimated by re-sampling the data 1,000 times. Panel {\it c} shows that the $\Ha$ extension has peak velocities that deviate from regular rotation, arguing against it being part of a skewed disk. 
\label{fig:kin}}
\end{figure*} 

Fig.~\ref{fig:spectra}d and Fig.~\ref{fig:spectra}e show the spectra centered on the $\Ha$ emission line of the highlighted spaxels in the $\Ha$ complex (which is $6.67\arcsec$ away from the center) and in the center of the host galaxy. Both spectra possess strong emission from the $\NII$ doublet, $\Ha$, and the $\SII$ doublet, confirming the authenticity of this $\Ha$ extension and indicating the presence of ionized gas throughout the entire system. Stacking all the spectra within the green circle centered in the $\Ha$ extension of Fig.~\ref{fig:spectra}b---which we will refer to as the ``$\Ha$ circle''---does not reveal a significant continuum. Properties of the host galaxy are listed in Table~\ref{tab:properties}.

Fig.~\ref{fig:logoh} presents the gaseous metallicities, 12+log(O/H), of this system using {\tt IZI} (Fig.~\ref{fig:logoh}a) and the $\NII/\OII$ calibration (Fig.~\ref{fig:logoh}b), with the $\Ha$ flux contours overlaid and the same spaxels from Fig.~\ref{fig:spectra}b highlighted in white. Using the estimates of {\tt IZI} (the $\NII/\OII$ calibration), the stacked 12+log(O/H) value in a $1.5\arcsec$ circle centered on the host galaxy is $9.06 \pm 0.06$ ($8.92 \pm 0.02$), whereas the stacked 12+log(O/H) value in the $\Ha$ circle is $8.81 \pm 0.05$ ($8.77\pm0.03$)---$0.25$ ($0.15$) dex less than the center of the host galaxy at greater than $99.7\%$ confidence. Comparing these gaseous metallicities to the solar value (12+log(O/H)$_{\odot}=8.69$; \citealt{asplund09}) reveals that the entire system has super-solar gaseous metallicities (but see \citealt{kennicutt03} for caveats associated with strong-line abundances). 

Fig.~\ref{fig:logoh}c and Fig.~\ref{fig:logoh}d display the 12+log(O/H) profile as traced by the highlighted spaxels in Fig.~\ref{fig:logoh}a and Fig.~\ref{fig:logoh}b, with the characteristic metallicity profile of non-interacting disks from \cite{sanchez14} overplotted in the dashed line (with an arbitrary zero-point). Out to $\sim2 R_{\rm e}$, the metallicity profiles of this system are similar to that of \cite{sanchez14}. At $\gs2 R_{\rm e}$, however, there appears to be a break that corresponds to the location of the $\Ha$ extension. 

The estimated metallicity profile of this system, however, is highly uncertain, as indicated by the large error bars from {\tt IZI}, which, unlike the $\NII/\OII$ calibration, considers multiple strong emission lines (see \S2). These large error bars from {\tt IZI} likely reflect the contrasting diagnostic line ratios, which may be due to the blending of physically different emission regions caused by insufficient spatial resolution. 

Another source of uncertainty is the DIG contribution to the emission line ratios, which could produce artificial metallicity gradients, especially in the outskirts of the system (Zhang et al., submitted). While we argue that DIG is not the major component of this system in \S\ref{sub:dig}, we cannot rule out minor DIG contributions that could affect the estimated metallicity gradients of this system.

The ionized gas kinematics of this system is presented in Fig.~\ref{fig:kin}, with $v_{\rm peak}$ displayed in panel a and $W_{\rm 80}$ in panel b; again, we superimpose the $\Ha$ flux contours and highlight the two spaxels from Fig.~\ref{fig:spectra}b. Fig.~\ref{fig:kin}a reveals an asymmetric gradient that ranges from $v_{\rm peak}\approx-140$ km s$^{-1}$ at the $\Ha$ extension to $v_{\rm peak}\approx40$ km s$^{-1}$ at the right side of the host galaxy. Fig.~\ref{fig:kin}b shows typical values of $W_{\rm 80}\approx230$ km s$^{-1}$, with areas of enhanced $W_{\rm 80}$ in the $\Ha$ extension and the outer envelope of the $\Ha$ flux distribution. The velocity profile of the highlighted spaxels in Fig.~\ref{fig:kin}a is shown in Fig.~\ref{fig:kin}c, with $R_{\rm e}$ of the host galaxy marked by the dashed, vertical lines. $|v_{\rm peak}|$ rises smoothly out to $\pm5\arcsec$, where it starts to flatten. Beyond $-5\arcsec$, however, there is an irregular wiggle in $v_{\rm peak}$ that corresponds to boosted $W_{\rm 80}$ in the $\Ha$ extension.

\begin{figure*}[t!] 
\centering
\includegraphics[scale=1]{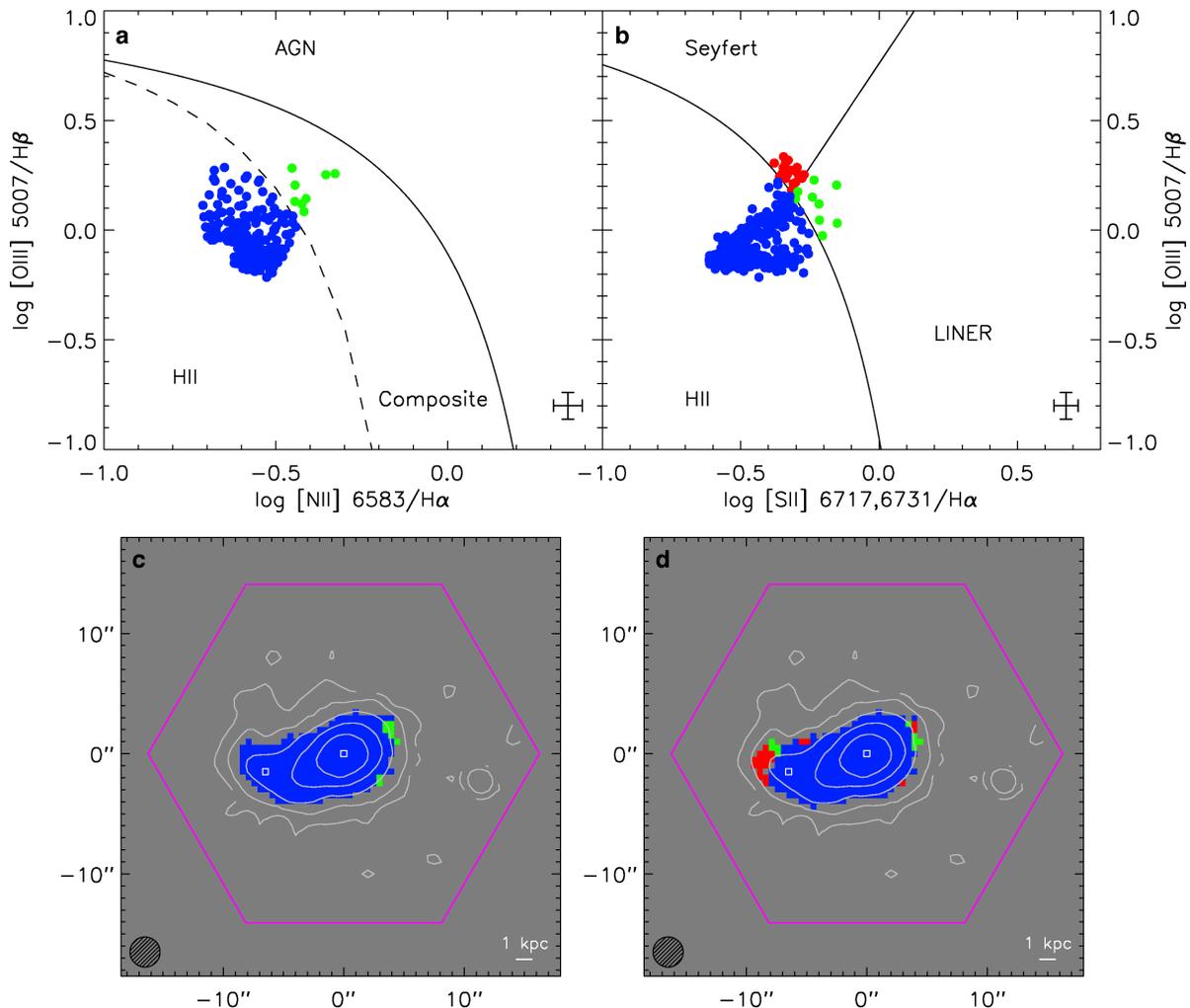}
\caption{ {\it a-b:} The \NII ~and \SII ~BPT diagrams; the lower right error bars represent the typical 1$\sigma$ measurement errors. {\it c-d:} The resolved \NII~and \SII~BPT diagrams, i.e., every spaxel is color-coded according to its location in the BPT diagrams above. Almost all the spaxels of this system have line ratios in the H{\sc ii} regime. 
\label{fig:bpt}}
\end{figure*}

The \NII~and \SII~BPT \citep{baldwin81} diagrams for this system are presented in Fig.~\ref{fig:bpt}a and Fig.~\ref{fig:bpt}b, respectively. We only consider spaxels with emission lines that have $S/N>3$ and overlay curves from \cite{kewley06} that are used to separate different classes of galaxies. Fig.~\ref{fig:bpt}c and Fig.~\ref{fig:bpt}d display the resolved \NII~and \SII~BPT diagrams, respectively; i.e., these maps color-code each spaxel according to its location in their respective BPT diagrams.  We overplot the $\Ha$ flux contours and highlight the same two spaxels in Fig.~\ref{fig:spectra}b. Fig.~\ref{fig:bpt} demonstrates that almost all spaxels have H{\sc ii} line ratios.

\section{Discussion} \label{sec:discussion}

In this section, we discuss the nature of the $\Ha$ extension.

\subsection{Diffuse ionized gas} \label{sub:dig}

\begin{figure*}[t!] 
\centering
\includegraphics[scale=1.17]{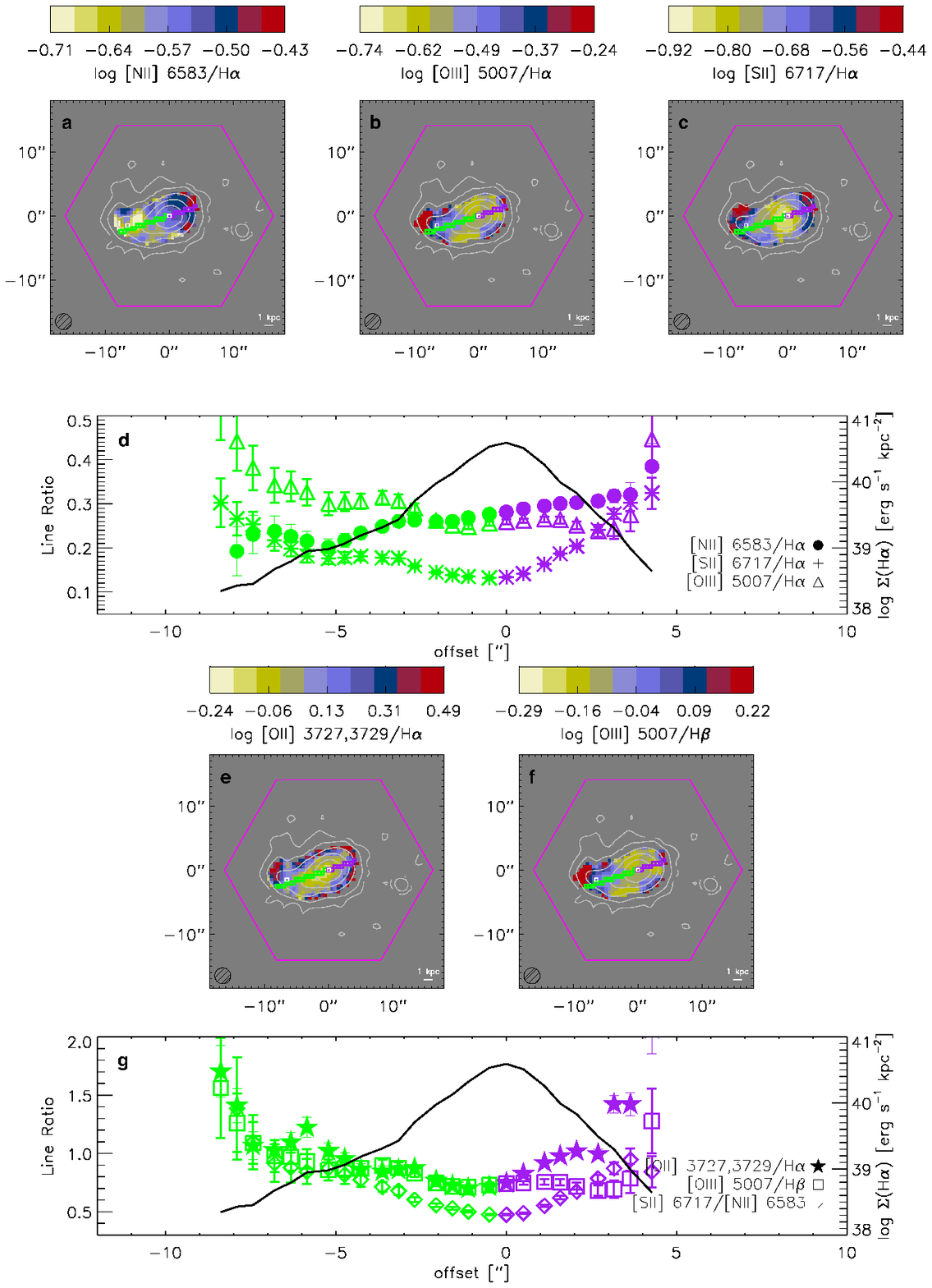}
\caption{ {\it a-c:} The $\log~\NII/\Ha$, $\log~\OIII/\Ha$, and $\log~\SII/\Ha$ line ratio maps. {\it d:} Line ratio profiles of $\NII/\Ha$, $\OIII/\Ha$, and $\SII/\Ha$ of the highlighted spaxels in panels {\it a-c}; the colors correspond to the colors of the highlighted spaxels and the error bars represent the 1$\sigma$ measurement error. The black line represents the $\Ha$ surface brightness profile across the same highlighted spaxels, with its values indicated by the y-axis on the right side. {\it e-f:} The $\log~\OII/\Ha$ and $\log~\OIII/\Hb$ line ratio maps. {\it g:} The line ratio profiles of $\OII/\Ha$, $\OIII/\Hb$, and $\SII/\NII$ of the highlighted spaxels in panels {\it e-f}. We only consider spaxels with $S/N>3$ for these emission lines and have corrected all these line ratios for reddening. 
\label{fig:line_ratio_profile}}
\end{figure*} 

\begin{figure}[t!] 
\centering
\includegraphics[scale=.75]{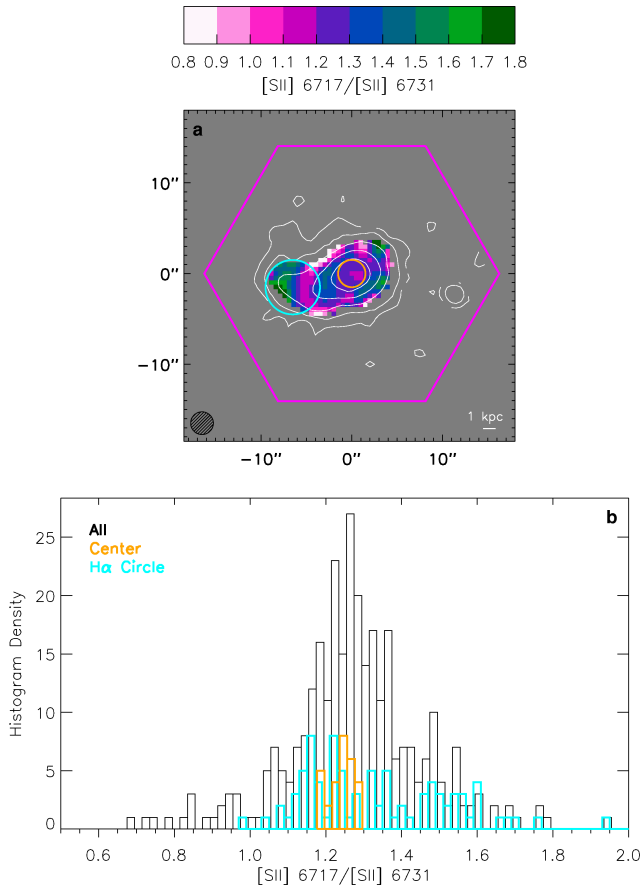}
\caption{ {\it a:} The $\SII~6717/\SII~6731$ line ratio map. {\it b:} The histogram of the $\SII~6717/\SII~6731$ line ratios of the entire system (black), the central region (orange, as indicated by the orange circle in the top panel), and the $\Ha$ circle (cyan).}
\label{fig:siiratio}
\end{figure}

The presence of warm, ionized gas in the outskirts of galaxies have been commonly referred to as diffuse ionized gas (DIG; \citealt{hoyle63, reynolds85}). The most pronounced characteristics of the DIG are the elevated $\NII/\Ha$ and $\SII/\Ha$ line ratios compared to H{\sc ii} regions \citep[e.g.,][]{monnet71,rand90, rand97, rand98, haffner99, tullmann00, otte01, otte02, collins01, hoopes03, wood04, voges06, rand08, haffner09, reynolds12}, which may be due to secondary ionization/heating sources \citep[e.g.,][]{reynolds92, reynolds99}. Thus comparing the line ratios of the $\Ha$ extension to that of H{\sc ii} regions would constrain whether the $\Ha$ extension is DIG. 

Ideally, we would identify individual H{\sc ii} regions in our system in order to conduct this comparison. However, the spatial resolution of the MaNGA data is too coarse---the FWHM is $2.5''$, which corresponds to $\approx1.9$ kpc at $z=0.038$ (the host galaxy's redshift), which is more than an order of magnitude larger than typical H{\sc ii} regions (1-100 pc; \citealt{kennicutt84, hunt09}). Therefore every spaxel in our data contains contributions from both H{\sc ii} regions and DIG.

One way to constrain the DIG contribution of the $\Ha$ extension is by comparing its line ratios to that of the center of the host galaxy, where the $\Ha$ surface brightness is the highest and the DIG contribution is the lowest \citep{ferguson96}. However, since line ratios are also dependent on metallicity and ionization parameter \citep{dopita00, dopita13, kewley02, kewley06}---which are likely to be different at the center compared to the $\Ha$ extension---another way to constrain the DIG contribution is to compare the line ratios of the $\Ha$ extension to the line ratios of the surrounding regions.

We present this line ratio comparison in Fig.~\ref{fig:line_ratio_profile}, all of which have been corrected for reddening according to \S\ref{sec:data}. Figs.~\ref{fig:line_ratio_profile}a-c show the log line ratio maps of $\NII/\Ha$, $\OIII/\Ha$, and $\SII/\Ha$, and Figs.~\ref{fig:line_ratio_profile}e-f show the log line ratio maps of $\OII/\Ha$ and $\OIII/\Hb$; the white contours represent the $\Ha$ flux distribution. Fig.~\ref{fig:line_ratio_profile}d displays the $\NII/\Ha$ (filled circles), $\OIII/\Ha$ (open triangles), and $\SII/\Ha$ (asterisks) line ratio profiles as traced by the highlighted spaxels in Fig.~\ref{fig:line_ratio_profile}a-c, while Fig.~\ref{fig:line_ratio_profile}g presents the $\OII/\Ha$ (filled stars), $\OIII/\Hb$ (open squares), and $\SII/\NII$ (open diamonds) line ratio profiles as traced by the highlighted spaxels in Fig.~\ref{fig:line_ratio_profile}e-f; the black solid line represents the $\Ha$ surface brightness profile over the same highlighted spaxels.

Fig.~\ref{fig:line_ratio_profile} indicates a complicated situation. While Fig.~\ref{fig:line_ratio_profile}d and Fig.~\ref{fig:line_ratio_profile}g show that $\NII/\Ha$ generally decreases with more negative offsets from the center of the galaxy, i.e., toward the direction of the $\Ha$ extension, the $\SII/\Ha$, $\OIII/\Ha$, $\OII/\Ha$, and $\OIII/\Hb$ generally increase along the same direction. These enhancing line ratios toward the $\Ha$ extension are consistent with DIG, but the decreasing $\NII/\Ha$ line ratio toward the $\Ha$ extension are not consistent with DIG  (e.g., \citealt{hoopes03, voges06}; Zhang et al. 2016, submitted). This latter trend is particularly striking since $\NII/\Ha$ is sensitive to the temperature of the gas and has been used to infer that the DIG is about $\sim2000$ K hotter than H{\sc ii} regions---a defining characteristic of DIG \citep{haffner99, madsen06, haffner09, reynolds12}. 

Fig.~\ref{fig:line_ratio_profile}d also shows that the $\NII/\Ha$ profile displays a dip around an offset of $-5''$, which corresponds to the location of the $\Ha$ extension, while the $\SII/\Ha$ profile and the $\Ha$ surface brightness profile display a flattening at the same spot. Since the regions surrounding the $\Ha$ extension should have a similar metallicity and ionization parameter, the fact that the $\Ha$ extension does not display enhanced $\NII/\Ha$ and $\SII/\Ha$ compared to its surrounding regions indicates that the $\Ha$ extension is not dominated by DIG. However, since the $S/N$ is low at the outskirts, the comparison of the line ratios at the center of the $\Ha$ extension (at $\sim-5''$) to those at the outermost regions are more uncertain.
  
Another defining characteristic of DIG is its low density ($\sim10^{-3}~\rm cm^{-3}$; e.g., \citealt{haffner09}). Hence another way to constrain the DIG contribution is to probe the gas density of the system with the \SII~6717/ \SII~6731 line ratio \citep{osterbrock06}. We present the \SII~6717/ \SII~6731 line ratio map in Fig.~\ref{fig:siiratio}a (with the $\Ha$ contours superimposed and the same spaxels from Fig.~\ref{fig:spectra}b highlighted in white), and the histogram of the \SII~6717/ \SII~6731 line ratios for various regions in Fig.~\ref{fig:siiratio}b.

The \SII~6717/ \SII~6731 line ratios of the $\Ha$ circle ranges from 1.0 to 1.6, which corresponds to electron densities, $n_{\rm e}$, of 500 cm$^{-3}$ to  $<10$ cm$^{-3}$, respectively \citep{proxauf14}. Even though the leftmost edge of the $\Ha$ circle shows electron densities that are less than 10 cm$^{-3}$, that a large area of the $\Ha$ extension contains dense gas with $n_{\rm e}\sim500 \rm~cm^{-3}$ strongly indicates that the $\Ha$ extension is not dominated by DIG. 

\subsection{Gas accretion} \label{sub:gas_accretion}

Alternatively, the $\Ha$ extension could be a sign of gas accretion \citep{sancisi08, rubin12, bouche13, almeida14, almeida15, bouche16}, which can manifest through a variety of ways, e.g., cold flows (the properties of this galaxy, $\log~M_*/\rm M_{\odot}=9.77$ and $z=0.038$, suggest cold flows rather than hot flows; \citealt{keres05, keres09a, dekel06, dekel09, danovich15, stewart16}), recycled wind material \citep{shapiro76, bregman80, fraternali08, oppenheimer08, keres09b, oppenheimer10, hopkins14}, or a gas-rich low-surface-brightness (LSB) dwarf galaxy \citep[e.g.,][]{leaman15, fischer15}.  

The super-solar gaseous metallicities of this $\Ha$ extension (see Fig.~\ref{fig:logoh}), however, is inconsistent with gas accretion through cold flows from the intergalactic medium (IGM) since this type of accreting gas is expected to be very metal-poor \citep{dave11, vandevoort12, joung12}. 
 
The H{\sc ii} line ratios of the $\Ha$ extension indicates that star-formation is the primary source of ionization (see Fig.~\ref{fig:bpt}), suggesting that the $\Ha$ extension is an accreting gas-rich LSB dwarf galaxy. Assuming that this $\Ha$ extension is indeed a LSB dwarf galaxy, then we estimate a stellar mass upper limit of $M_*\sim10^{7}~\rm M_{\odot}$.

To obtain this limit, we first sum the $g$ band flux of every pixel in the $\Ha$ circle using SDSS DR7; we multiply this sum by three and take it as the upper limit of the $g$ band flux. Selecting SDSS DR7 galaxies at $0.02<z<0.04$, we plot their mass-to-light ratios ($M/L$) as a function of rest-frame $g-r$ color using the MPA-JHU stellar masses and galaxy $g$ and $r$ band absolute magnitudes from GIM2D \citep{simard11}, and then use the typical $M/L$ of star-forming galaxies ($0.2<g-r<0.4$) to estimate the stellar mass upper limit of the $\Ha$ extension, which yields $M_*\sim10^{7}~\rm M_{\odot}$.

A galaxy with such a low stellar mass, however, is unlikely to have the super-solar gaseous metallicities that are observed in the $\Ha$ extension \citep{berg12, jimmy15}, indicating that the $\Ha$ complex is probably not an accreting gas-rich LSB dwarf galaxy. 

Finally, the enhanced values of $W_{\rm 80}$ within the $\Ha$ extension (see Fig.~\ref{fig:kin}b) that coincide with an unusual $v_{\rm peak}$ bump in the velocity profile (at $\sim-7\arcsec$ of Fig.~\ref{fig:kin}c) may suggest the presence of turbulence that could be caused by the interaction between accreting gas and the host galaxy. However, such large values of $W_{\rm 80}$ may also be an artifact caused by the coarse resolution of the MaNGA data, which could blend multiple components separated by small velocities and/or small distances. Future higher resolution data will be needed to discern these possibilities.

Therefore the gas accretion scenario that best matches the properties of the $\Ha$ extension is recycled wind material \citep{oppenheimer08, keres09b, oppenheimer10, hopkins14}, which is predicted to be a critical process in shaping the galaxy stellar mass function at low stellar masses ($<5\times10^{10} ~M_*/\rm M_{\odot}$; \citealt{oppenheimer10, hopkins14}). 

\subsubsection{Gas accretion deficit} \label{sub:gas_accretion_deficit}

If this $\Ha$ extension is indeed accreting gas, then an interesting application would be toward the gas accretion deficit. To elaborate, \cite{sancisi08} combined many nearby H{\sc i} studies to show that the visible H{\sc i} gas accretion rate in the local universe is only $\sim 0.2~$ \rm M$_{\sun}$ yr$^{-1}$, which is about an order of magnitude too small to account for the current $SFR$ of local star-forming galaxies ($\sim1$ M$_\sun$ yr$^{-1}$). These H{\sc i} studies, however, only probe cold gas $(T\ls100$ K). Simulations predict that most of the gas in cold flows have $T\gs10^4$ K upon entering the host galaxy \citep{vandevoort12, joung12, nelson13}. Thus probing warm ($\sim10^4$ K), ionized gas from the CGM may alleviate the gas accretion deficit.

To test this idea, we compare the potential warm gas accretion rate onto this galaxy with its $SFR$. We first calculate the ionized gas mass of the $\Ha$ circle using the following equation from \cite{osterbrock06}:

\begin{equation} \label{eqn:ion_mass}
\frac{M_{\rm ionized~gas}}{2.82\times10^9 ~\rm M_{\sun}} = \Bigg( \frac{L_{\rm H\beta}}{10^{43}~\rm erg~s^{-1}} \Bigg) \Bigg( \frac{n_{\rm e}}{100~\rm cm^{-3}} \Bigg)^{-1} ,
\end{equation}

\noindent where $L_{\rm H\beta}$ is the extinction-corrected $\Hb$ luminosity and $n_{\rm e}$ is the electron density. We calculate $L_{\rm H\beta}$ by summing the extinction-corrected $\Hb$ luminosities of every spaxel in the $\Ha$ circle, yielding $L_{\rm H\beta}=2.89\times10^{39}~\rm erg~s^{-1}$. To estimate $n_{\rm e}$, we measure the median $\SII$~6717/$\SII$~6731 ratio ($\approx1.3$; see Fig.~\ref{fig:siiratio}b) of the $\Ha$ circle to obtain $n_{\rm e}\approx50~\rm cm^{-3}$; this calculation yields $M_{\rm ionized~gas} \approx 1.63 \times 10^6 ~\rm M_{\sun}$.

The time required for this $\Ha$ circle to accrete onto the host galaxy, $t_{\rm accretion}$, is estimated by dividing the distance between the $\Ha$ circle and the host galaxy with the observed velocity of the $\Ha$ circle. This distance, i.e., the distance between the two highlighted spaxels in Fig.~\ref{fig:spectra}b, is $\approx6.67 \arcsec$, which corresponds to $1.5\times10^{17}$ km. The median $|  v_{\rm peak} | $ of the $\Ha$ circle is $\approx 80$ km s$^{-1}$. Hence $t_{\rm accretion}\approx 6\times10^{7}$ yrs. 

Therefore, the average warm gas accretion rate onto the host galaxy is $M_{\rm ionized~gas}/t_{\rm accretion}\approx0.03~\rm M_{\sun}~\rm yr^{-1}$. If this accreting warm gas were driving the current star-formation in the host galaxy, we would expect for this warm gas accretion rate to be similar to the current $SFR$. However, the estimated star-formation rate of the host galaxy is $SFR=1.67~\rm M_{\rm \sun}~yr^{-1}$, which is almost two orders of magnitude larger than the estimated warm gas accretion rate. Thus warm gas accretion does not appear to solve the gas accretion deficit for this galaxy. 

However, since this galaxy is starbursting, which is defined to be a short period of intense star-formation, perhaps it is unsurprising that the warm gas accretion does not sustain the current star-formation rate. Moreover, this estimated warm gas accretion rate may be a lower limit since we do not account for transverse motions, meaning the true velocity could be higher, nor do we account for the possibility of more diffuse ionized gas that is below our detection limit. 

\subsection{Other possibilities}

Another possibility for this $\Ha$ complex is that it's an outflow. If it were an outflow from a starburst though, we'd expect for its 12+log(O/H) values to be the same as, or higher than, the center of the host galaxy, where the metal-enriching stellar feedback would have likely originated. Instead, we find that the $\Ha$ extension has lower 12+log(O/H), arguing against an outflow. Another possibility is that it's gas disrupted and/or stripped by environmental processes, e.g., mergers or ram-pressure stripping \citep{gunn72}. However, since the host galaxy shows no signs of interaction and does not appear to be in a large group, we find this explanation unlikely. Moreover, with a galaxy S\'ersic index of $n=3.7$, unusual gaseous metallicity profile, and irregular rotation curve, it is also unlikely that this $\Ha$ extension is simply part of the host galaxy as a dim, lopsided disk.  

Finally, the elongated, asymmetric $\Ha$ flux distribution of the system (see Fig.~\ref{fig:spectra}) resembles the morphologies of tadpole galaxies \citep{vandenbergh96, elmegreen05, elmegreen07, elmegreen12, almeida13, straughn15}, suggesting that this $\Ha$ extension may be part of a tadpole galaxy. However, in addition to their cometary morphology, another common property of tadpole galaxies is their bright, blue colors that are indicative of recent bursts of intense star formation \citep{elmegreen10, elmegreen12}. Since the $\Ha$ extension in our system has no optical or ultraviolet (UV) counterpart, it is unlikely that our system is a tadpole galaxy. Moreover, the super-solar gaseous metallicities of our system is inconsistent with the very metal-poor gaseous metallicities found in local tadpole galaxies \citep{almeida13}. 

\section{Conclusion}

In this paper, we present the serendipitous observation of an ionized gas structure that protrudes out of a dusty, starbursting galaxy. Our analysis indicates that this ionized gas complex is most consistent with gas accretion through recycled wind material.

To better understand the nature of this phenomenon, and to constrain the importance of gas accretion through recycled wind material, we need to find more of these extended gas complexes. However, the current MaNGA sample has not observed any other isolated galaxy with an adequately-sized integral field unit ($\gs6.3R_{\rm e}$) to allow such a search. We hope to address this issue with future MaNGA releases.

\acknowledgments

We thank Kevin Bundy, Kyle Westfall, and Matthew Bershady for helpful comments and discussions. We also thank the anonymous referee for a constructive report that improved this work. 

Funding for the Sloan Digital Sky Survey IV has been provided by
the Alfred P. Sloan Foundation, the U.S. Department of Energy Office of
Science, and the Participating Institutions. SDSS-IV acknowledges
support and resources from the Center for High-Performance Computing at
the University of Utah. The SDSS web site is www.sdss.org. 
SDSS-IV is managed by the Astrophysical Research Consortium for the 
Participating Institutions of the SDSS Collaboration including the 
Brazilian Participation Group, the Carnegie Institution for Science, 
Carnegie Mellon University, the Chilean Participation Group, the French Participation Group, Harvard-Smithsonian Center for Astrophysics, 
Instituto de Astrof\'isica de Canarias, The Johns Hopkins University, 
Kavli Institute for the Physics and Mathematics of the Universe (IPMU) / 
University of Tokyo, Lawrence Berkeley National Laboratory, 
Leibniz Institut f\"ur Astrophysik Potsdam (AIP),  
Max-Planck-Institut f\"ur Astronomie (MPIA Heidelberg), 
Max-Planck-Institut f\"ur Astrophysik (MPA Garching), 
Max-Planck-Institut f\"ur Extraterrestrische Physik (MPE), 
National Astronomical Observatory of China, New Mexico State University, 
New York University, University of Notre Dame, 
Observat\'ario Nacional / MCTI, The Ohio State University, 
Pennsylvania State University, Shanghai Astronomical Observatory, 
United Kingdom Participation Group,
Universidad Nacional Aut\'onoma de M\'exico, University of Arizona, 
University of Colorado Boulder, University of Oxford, University of Portsmouth, 
University of Utah, University of Virginia, University of Washington, University of Wisconsin, 
Vanderbilt University, and Yale University.  
DB is supported by grant RSCF-14-22-00041. AW acknowledges support from a Leverhulme Early Career Fellowship. 
JHK acknowledges financial support from the Spanish Ministry of Economy and Competitiveness (MINECO) under grant number AYA2013-41243-P, and thanks the Astrophysics Research Institute of Liverpool John Moores University for their hospitality, and the Spanish Ministry of Education, Culture and Sports for financial support of his visit there, through grant number PR2015-00512.


\end{document}